\documentclass[11pt,a4paper]{article}
\usepackage{jheppub}
\usepackage{graphicx,color}
\usepackage[latin1]{inputenc}
\usepackage{amsmath,amssymb}
\usepackage{hyperref}
\usepackage{slashed}
\newcommand{\specialcell}[2][c]{%
  \begin{tabular}[#1]{@{}c@{}}#2\end{tabular}}

\title{A Scenario of Heavy but Visible Baryonic Dark Matter}
\author{Ran Huo, Shigeki Matsumoto, Yue-Lin Sming Tsai, Tsutomu T. Yanagida}
\affiliation{Kavli IPMU (WPI), UTIAS, The University of Tokyo, Kashiwa, Chiba 277-8583, Japan}

\emailAdd{ran.huo@ipmu.jp}
\emailAdd{shigeki.matsumoto@ipmu.jp}
\emailAdd{yue-lin.tsai@ipmu.jp}
\emailAdd{tsutomu.tyanagida@ipmu.jp}

\abstract{
We consider a model in which dark matter is a composite baryon of a dark sector governed by $SU(3)$ gauge theory, with vector-like quarks also charged under $U(1)_Y$. The model provides simple answer to the dark matter stability problem: it is a result of the accidental dark baryon number conservation. And with an analogy to QCD, all physical quantities of the dark matter can be calculated by rescaling the QCD experimental results. According to the thermal freeze-out mechanism the mass of the dark matter is predicted to be $\mathcal{O}(100)$~TeV in order to achieve a correct relic abundance. Such heavy dark matter is in general hard for detection due to small dark matter number density in the universe. However, dark baryon number in our model is not necessarily strictly preserved thanks to operators suppressed by the Planck scale, and such decay operator results in a decay lifetime marginal to the current detection bound. We show our model with $\mathcal{O}(10^{27})~s$ dark matter decay life time can explain the AMS-02 anti-proton data, if it is experimentally interpreted as an access, although some theoretical uncertainty may weaken its significance. We also investigate other phenomena of this model such as the extragalactic gamma ray and neutrino signatures.
}
\begin{document}
\maketitle

%%%%%%%%%%%%%%%%%%%%%
%%%%% Introduction %%%%%
%%%%%%%%%%%%%%%%%%%%%
\section{Introduction}

A lot of particle physics candidates have been worked out for dark matter (DM), which takes about $26\%$ of the energy density of our universe~\cite{Ade:2015xua}. All of the models have, at least after tuning to certain parameters, to explain some basic facts of DM such as its stability, weakness of its interaction with the standard model (SM) matter, its relic abundance and so on. Mainly motivated by the stability problem, we revisit the composite DM model, in which DM is a baryonic or antibaryonic composite particle in a hidden strong gauge interaction, or the technibaryon DM which has a long history~\cite{Nussinov:1985xr,Barr:1990ca,Chivukula:1989qb} (see more recent similar works~\cite{Gudnason:2006yj,Hambye:2009fg,Hamaguchi:2009db,Murayama:2009nj,Antipin:2014qva,Antipin:2015xia}, and more generally the references in~\cite{Antipin:2015xia}).

In the simplest put our composite DM model is a copy of quantum chromodynamics (QCD) at different scale. Some hidden strong gauge group at certain scales has exactly the same QCD phenomena of color confinement and chiral symmetry breaking, they contain fermionic dark baryons as composite bound states in totally antisymmetric representation. In particular we will see that our model also has an $SU(3)$ strong gauge group, which will be introduced below. Protected by the accidental baryon number, the QCD proton is stable empirically, even if the theory is promoted to the grand unified theory (GUT) and proton decay is allowed. The stability of (the lightest) dark baryon is similarly guaranteed, as a consequence of the accidental dark baryon number conservation.

Within baryonic DM models there are still a lot of possibilities, and some of the models are actually difficult to test. If the dark sector and the SM sector are connected only by some portal particle, then the typical interaction between the two sectors may be too week to see (e.g. the direct detection cross section). Moreover, for reproducing the correct relic abundance we would like to stick to the simplest assumptions that the DM is a thermal relic (as well as implicitly the dark baryon antibaryon are symmetric in number). However, in the following we will see that the repetition of the QCD exactly provides us a tool to calculate the thermal freeze out cross section, and the DM mass is therefore determined to be $\mathcal{O}(100)$~TeV~\cite{Hambye:2009fg,Hamaguchi:2009db,Murayama:2009nj,Antipin:2014qva,Antipin:2015xia}. Such high scale DM mass implies a very small DM number density in the universe, and suppressing any DM event rates such as direct detection and annihilation, making it invisible. In particular, we have checked that even with the largest possible annihilation cross section of the $s$ wave unitarity bound, the most optimistic NFW DM profile and large exposure time assumptions, next generation gamma ray indirect detection experiment such as the CTA still cannot probe/constrain such a high scale DM candidate from the galaxy center, similar to~\cite{Ibarra:2015tya}.

Still we manage to test our model. Instead of connecting the two sectors with a portal, we consider the following criteria~\cite{Hamaguchi:2009db} that, while the dark baryon itself must carry no net SM charges for being sufficiently weakly interacting with the visible sector, the vector-like quarks in the hidden sector are charged under some SM gauge groups of $SU(3)_c\times SU(2)_L\times U(1)_Y$. And the second problem of being too heavy on the other hand weakens the DM stability if it is not absolute, eventually enhance the decay rate to a detectable level as a benefit. For our benchmark $SU(3)_\text{hid}\times SU(2)_R\times U(1)_{B-L}$ model the decay spectrum can be determined and the decay rate happens to be marginal to the current bound, which eventually makes the model testable. %\footnote{For the whole scenario to work a few assumptions about the scales are needed, which will be stated below.}.
For example we compare our model prediction of DM decay against the preliminary AMS-02 antiproton data, as well as checking with the Fermi extragalactic gamma ray and the IceCube neutrino data. In all, (unlike many early work) we have provided answers to every aspect of DM physics, by building a model which is simple, and detectable or falsifiable in the near future.

This paper is structured as followed. Section~\ref{sec:stab} is a quick check of the stability in more details. In Section~\ref{sec:ann} we discuss the rescaling calculation of the DM annihilation cross section, and eventually show the DM mass scale of $150$~TeV. Next Section~\ref{sec:mod} is devoted to the discussion of the specific gauge group choice, in which an $O(10^{27})$~s decay life time is estimated and the $SU(3)_\text{hid}\times SU(2)_R\times U(1)_{B-L}$ is justified, with discussion of constraints (such as the one from the big bang nucleosynthesis) for various scales implicitly in the model building. We check the recent AMS-02 antiproton data with the decay life time as a major test of the model in Section~\ref{sec:dec}, as well as check other constraints such as the extragalactic gamma ray and neutrino flux. At last we conclude in Section~\ref{sec:sum}.

%%%%%%%%%%%%%%%%%%%%%
%%%%% Introduction %%%%%
%%%%%%%%%%%%%%%%%%%%%
\section{The Dark Matter Stability\label{sec:stab}}

The solution of DM stability, or suppression of any possible decay operators by compositeness, still allows a range of hidden gauge groups. Suppose the hidden strong gauge group is an $SU(N)$ gauge group, with $N$ an odd number for baryon to be a fermion. The dark baryon is in the totally antisymmetric representation of $N$ dark quarks, and as the proton decay in GUT the effective decay operator is at least $N+1$ fermions times together, with the new ``$1$'' being some (not necessarily to be new, as we will see) ``lepton''. To be overall mass dimension 4 we need high scale $\Lambda$ suppression of power $\frac{3}{2}(N+1)-4$.
In such analysis the DM has an decay width of order
\begin{equation}
\Gamma\sim m_\chi^{3N-4}/\Lambda^{3N-5}.\label{eq:decay}
\end{equation}
This hidden strong gauge group has no relation with the postulated GUT in the visible sector, so the first choice of the large scale is the Planck scale. Then we can see that even for a minimal $N=3$, the DM can be stable enough. However, the choice of $N=3$ will actually turn out to be tricky and interesting, as shown in Section~\ref{sec:mod}.

Further comment is that we expect this dimension-6 $QQQL$ operator for $N=3$ gives leading contribution to DM decay, simply because given the dimension of $3\times\frac{3}{2}$ (with $\frac{3}{2}$ being the quark dimension) which must appear as a whole in any possible decay operator, it is unable to construct any operator with even lower dimension. Within the framework of dimension-6 $QQQL$ operator the decay process exactly resembles the QCD proton decay, and the lattice calculation of the QCD proton decay matrix element can be rescaled to give results in the dark sector, given that the lepton $L$ mass are much lower. Specific choice such as whether the $L$ is charged or neutral will depend on the model building, which will be addressed in Section~\ref{sec:mod}. Apparently going to even higher dimensions there are various operator choices such as just adding powers of the SM $H^\dag H$ to the operator, but they will have much more suppressed decay contributions and are irrelevant.

%%%%%%%%%%%%%%%%%%%%%%%%%%%%%%%
%%%%%%%%%%% Annihilation %%%%%%%%%%%
%%%%%%%%%%%%%%%%%%%%%%%%%%%%%%%
\section{Dark Matter Annihilation and Mass\label{sec:ann}}

Due to the strong dynamics nature there is no way to calculate from the first principle the annihilation cross section of baryons with antibaryons at the low velocity regime. A useful first approximation is given by the unitarity bound~\cite{Griest:1989wd}
\begin{equation}
\sigma v_\text{rel}\simeq \frac{4\pi(2\ell+1)}{m_B^2v_\text{rel}}.\label{eq:unitarity}
\end{equation}
This unitarity bound can be obtained by solving a nonrelativistic Schrodinger equation, with a complex potential to decrease the exiting wave amplitude. Achieving the unitarity bound corresponds to a vanishing exiting $\ell$ partial wave, which is a result of alignment of the potential configuration. Potential scattering is nonperturbative in nature, in the Feynman diagram point of view messenger particles are repeatedly exchanged, forming ladder diagrams. And enhancement of the hard scattering/annihilation process are taken into account in this way, recovering the Sommerfeld enhancement~\cite{Hisano:2003ec,Hisano:2004ds}.

%%%%%%%%%%%%%%%%%%%%%%%%%%%%%%%%%%%%%%%%%%%%%%%%%%%%%%%%%%
\begin{figure}[t!]
\centering
\includegraphics[width=3in]{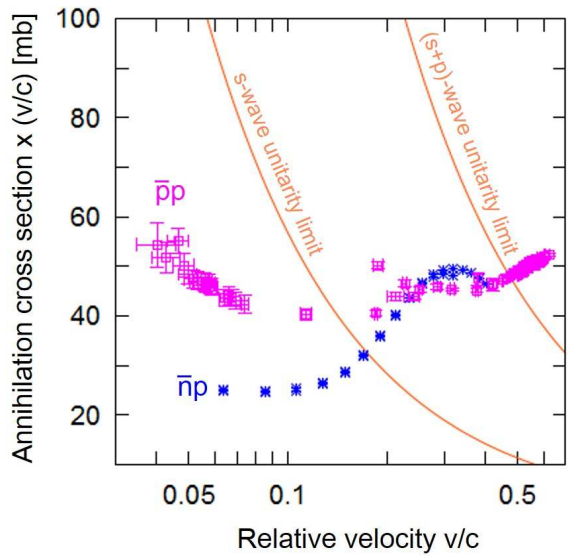}
\caption{The QCD measurement of baryon antibaryon annihilation cross section $\sigma v$ in the low relative velocity region. At higher relative velocity $\sigma v$ will exceed the $s$ wave or the $(s+p)$ wave unitarity bound. $\bar{p}p$ data are taken from~\cite{Bruckner:1989ew,Bertin:1996kw,Benedettini:1997fk,Zenoni:1999st}, and $\bar{n}p$ data are taken from~\cite{Armstrong:1987nu,Bertin:1997gn}.}
\label{fig:freeze}
\end{figure}
%%%%%%%%%%%%%%%%%%%%%%%%%%%%%%%%%%%%%%%%%%%%%%%%%%%%%%%%%%
The approximation of Eq.~(\ref{eq:unitarity}) is indeed supported by the real QCD experiments. In the low velocity regime higher partial waves are suppressed by powers of velocity, leading to a general $s$ wave domination if it is not forbidden by symmetries. However for a thermal freeze out velocity of $v_\text{rel}\simeq0.4$, the $s$ wave may be further assisted by a $p$ wave or even higher partial waves. In Fig.~\ref{fig:freeze} we see from experimental results that the annihilation cross section is indeed higher than $s$ wave unitarity bound by a factor of a few, due to mainly the $p$ wave contributions. A significant $p$ wave contribution same to our case has already been shown in~\cite{Bertin:1997gn}.

In Eq.~(\ref{eq:unitarity}) the true QCD annihilation cross section scales as $m_B^{-2}$ with $m_B$ the QCD baryon mass, while the $2\ell+1$ adopt an effective value of about two to four (note that the annihilation remains rather constant with variation of relativity velocity in Eq.~(\ref{eq:unitarity}), which means the effective $2\ell+1$ has a rough scaling with $v_\text{rel}$, to cancel the $v_\text{rel}$ dependence). If it is applied to the dark sector baryon antibaryon annihilation the same effective $2\ell+1$ should be kept, difference is only in replacing the QCD baryon mass (about $1~\text{GeV}$) by the dark baryon/antibaryon mass. Namely the DM annihilation cross section is just a rescaling of the QCD annihilation cross section
\begin{equation}
\sigma v_\text{rel}^\texttt{DM}(m_\chi)=\sigma v_\text{rel}^\texttt{QCD}\times \left(\frac{1~\text{GeV}}{m_\chi}\right)^2,
\label{eq:RtoDM}
\end{equation}
where $\sigma v_\text{rel}^\texttt{QCD}$ is the averaged value of purple and blue points in the Fig.~\ref{fig:freeze}. In the following we will see that our benchmark model is also based on the $SU(3)$ strong gauge group; and we are taking implicitly that the current quark mass is at least one order smaller than the baryon mass, making the chiral symmetry breaking effect small as the QCD. In such case the effective theory can be approximated by only one scale, which is the QCD scale of typical baryon mass for the real QCD, and the dark baryon mass for the dark sector\footnote{The other scales introduced in Section~4.2,4.3 has nothing to do with the annihilation.}. Thus the strong dynamics of the dark sector is quite similar to the real QCD case, and we expect the rescaling calculation to be very good.

The thermal freeze out mechanism is the most elegant way to reproduce the observed relic abundance, in which we only set the DM annihilation cross section to the famous $3\times10^{-26}~\text{cm}^3\thinspace\text{s}^{-1}$, corresponding to $\Omega_\chi h^2\simeq0.12$. When we do this to the baryonic dark sector we get
\begin{equation}
m_\chi\simeq150~\text{TeV}.\label{eq:mass}
\end{equation}

Here we make a few comments and comparisons on the result, which is crucial to the following analysis. This number is larger than the literature value of the $s$ wave unitarity bound of the DM mass in thermal freeze out, which calculated from Eq.~(12) of~\cite{Griest:1989wd} is about $86$~TeV for a Dirac fermion. Note that the latter number is assuming a freeze out relative velocity of about $0.4c$. In Fig.~\ref{fig:freeze} at such relative velocity we can read the true baryonic annihilation cross section is almost four times larger than the $s$ wave unitarity bound, so a further factor of two difference of DM mass can be understood. The discussion for higher partial waves in~\cite{Griest:1989wd} is implicitly assuming that one partial wave will dominate the total annihilation cross section. This assumption fails for a strong dynamics system such as the QCD case, which has already been shown in Fig.~\ref{fig:freeze} for the real QCD. Still in the same small relative velocity regime due to the strong dynamics nature many partial waves contribute to the annihilation amplitude and possibly interfere with each other, in that case the implicitly omitted $2\ell+1$ factor in~\cite{Griest:1989wd} makes a difference.

%%%%%%%%%%%%%%%%%%%%%%%%%%%%%%%
%%%%%%%%%%% Model %%%%%%%%%%%
%%%%%%%%%%%%%%%%%%%%%%%%%%%%%%%
\section{The Benchmark $SU(3)_\text{hid}\times SU(2)_R\times U(1)_{B-L}$ Dark Matter Model\label{sec:mod}}

\subsection{The Hidden Gauge Group}

The high scale of DM does provide one advantage, that the increase of decay width from Eq.~(\ref{eq:decay}) makes better chance of its detection in the decay channel. We start with checking $N=3$, a reduced Planck scale\footnote{Here we use the reduced Planck scale $(8\pi G)^{-\frac{1}{2}}=2.4\times10^{18}$~GeV rather than the Planck scale $G^{-\frac{1}{2}}=1.2\times10^{19}$~GeV. The reduced Planck scale appears naturally in supergravity theory of particle physics. The Planck scale, on the other hand, already gives too low a decay rate.} square suppression and the dark baryon mass of $150$~TeV. Eq.~(\ref{eq:decay}) gives $\Gamma\simeq m_\chi^5/\Lambda_\text{Pl}^4\approx(150\times10^3)^5/(2.4\times10^{18})^4=2.3\times10^{-48}$~GeV, which corresponds to a decay life time of $2.9\times10^{23}$~s. This estimation has already revealed an amazing coincidence, that the minimal model leads to a decay life time close to the current observational limit.

We can use lattice calculation of the GUT proton decay to further improve the estimation. Rescaling Eq.~(10) of~\cite{Aoki:2013yxa} of such calculation gives further significant corrections of an extra factor of $1/(32\pi)$ from the phase space, and another $10^{-2}$ or so as the lattice calculated matrix element amplitude square of $W_0^2\simeq(0.1~\text{GeV}^2)^2$ divided by the QCD proton mass to the fourth. Putting altogether the decay life time is a few times $10^{27}$~s.

The choice of hidden gauge group of $SU(3)$ has additional benefit, of further justifying the previous rescaling of the annihilation calculation. The real QCD has strong gauge group of $SU(3)_c$, or baryon consisting three quarks. The choice of $SU(3)_\text{hid}$ also preserves this structure, which is expected to minimize the uncertainty of our ignorance of strong dynamics, making the rescaling the most reliable compared with other choices of the strong gauge group.

\subsection{Introduction of the $SU(2)_R\times U(1)_{B-L}$}

Let's take a look at other aspects of the minimal $SU(3)_\text{hid}$ model. The requirement that it is also charged under some SM gauge group is minimally satisfied for the SM $U(1)_Y$ group, and that will induce the SM electromagnetic interactions and the current quarks will be SM electromagnetically charged. To further build our model we refer to the most common way of extension of the $U(1)_Y$ in beyond SM model building, namely the $SU(2)_R\times U(1)_{B-L}\to U(1)_Y$ left-right symmetry model, which also gives the neutrino masses. An $SU(2)_R$ gauge group behaves as the counterpart of the $SU(2)_L$ isospin, and the $U(1)$ before this breaking becomes $B-L$. The hidden quarks are postulated to be in the doublet representation of $SU(2)_R$.

\begin{table}[!htb]
\centering
\begin{tabular}{c c c c c c c}
 \hline\hline
 & Gauge & $SU(3)_H$ & $SU(3)_c$ & $SU(2)_R$ & $SU(2)_L$ & $U(1)_{B-L}$ \\
 \hline
New Particle & \specialcell{$\Phi$ \\ $Q_L$ \\ $Q_R$} & \specialcell{1 \\ 3 \\ 3 } & \specialcell{1 \\ 1 \\ 1 }
& \specialcell{3 \\ 2 \\ 2 } & \specialcell{1 \\ 1 \\ 1 } & \specialcell{$+1$ \\ $+\frac{1}{6}$ \\ $+\frac{1}{6}$ } \\
 \hline
SM & \specialcell{$H$ \\ $q_L$ \\ $q_R$ \\ $l_L$ \\ $l_R$ } & \specialcell{1 \\ 1 \\ 1 \\ 1 \\ 1 } & \specialcell{1 \\ 3 \\ 3 \\ 1 \\ 1 }
& \specialcell{2 \\ 1 \\ 2 \\ 1 \\ 2 } & \specialcell{2 \\ 2 \\ 1 \\ 2 \\ 1 } & \specialcell{0 \\ $+\frac{1}{6}$ \\ $+\frac{1}{6}$ \\ $-\frac{1}{2}$ \\ $-\frac{1}{2}$ } \\
 \hline\hline
\end{tabular}
\caption{The particle content and their quantum numbers.}
\label{tab:quantum_number}
\end{table}

Going back to the minimal dimension-6 decay operator $Q_RQ_RQ_R\ell_R$, the ``lepton'' is also in the doublet representation of $SU(2)_R$. Actually it can be identified as the SM right handed charged lepton, which is also promoted by the beyond SM $SU(2)_R$. The model is shown in Table~\ref{tab:quantum_number}. Here we also introduce an $SU(2)_R$ triplet $\Phi$, which gets some large vacuum expectation value to break the $SU(2)_R$, and gives large masses to the $SU(2)_R$ $W$ boson and the right handed neutrino~\cite{neutrinobook}. The right handed neutrino mass scale is postulated to be determined by the natural seesaw mechanism, or much larger than the DM scale of $150$~TeV, so the $\ell_R$ could only be the charged right handed lepton, in order not to be kinetically forbidden. We do not expect forbidden of right handed neutrino decay channels significantly affects the estimation of the DM life time.

At the level of the elementary particle our model can be summarized as the renormalizable Lagrangian
\begin{align}
\mathcal{L}&\supset\mathcal{L}_\text{LR}-(\lambda\ell_R^T\epsilon\Phi\ell_R+h.c.)\nonumber\\
&-\frac{1}{4}G_{\mu\nu}^aG^{a\mu\nu}-\frac{1}{4}W_{\mu\nu}^aW^{a\mu\nu}-\frac{1}{4}F_{\mu\nu}F^{\mu\nu}\nonumber\\
&+\bar{Q}_L(i\slashed{D}-M)Q_L+\bar{Q}_R(i\slashed{D}-M)Q_R+(D_\mu\Phi)^\dag D^\mu\Phi.
\label{eq:Lagrangian}
\end{align}
The $\mathcal{L}_\text{LR}$ is the left-right symmetric extension of the SM, and we write the allowed neutrino Majorana mass terms separately. Note the $\epsilon=i\sigma^2=(_{-1}~^{1})$ in the $SU(2)$ contraction, and in this matrix basis $\langle\Phi\rangle=(_V^0~^0_0)$. Each covariant derivative can be read from Table~\ref{tab:quantum_number}.

Note that if we choose the SM $SU(2)_L$ rather than the $SU(2)_R$ for the dark quark, then the dark baryon is also charged under the SM $SU(2)_L$. This still suits the definition of ``WIMP'', but the scattering with ordinary matter in direct detection will have charge induced $W^\pm$ scattering rather than the current dipole induced photon scattering as the leading one, and we do not consider this situation in our scenario.

\subsection{The Mass Splitting between Dark Neutron and Proton}

After the introduction of $SU(2)_R$, the baryonic DM candidate is also promoted into doublet. The dark neutron is the DM candidate. The dark proton need to be heavier than the dark neutron, and after thermal freeze out decay early enough to it.

Dimensional analysis suggests such decay has a width of $\Gamma\propto\Delta m^5/m_W^4$~\cite{Wilkinson:1982hu}, where $m_W$ is the SM (or $SU(2)_R$) $W$ boson mass for real QCD (or the dark sector), and $\Delta m$ is the mass difference between the SM (or dark) proton and neutron. %(We have ignored the possibility of producing an on-shell light dark pion.)
The simple requirement that the dark proton to dark neutron decay life time is before the big bang nucleosynthesis ($\lesssim1$~s) corresponds to a constraint of
\begin{equation}
\Delta m\gtrsim0.22\left(\frac{m_{W_R}}{10^8~\text{GeV}}\right)^{\frac{4}{5}}~\text{TeV},\label{eq:splitting}
\end{equation}
which is obtained by rescaling the QCD free neutron to proton decay.

The mass difference between the dark proton and the dark neutron comes in two ways: One is the electromagnetic radiative self energy correction which applies only to charged proton. The other is the current quark mass differences. In real QCD the two happen to cancel with each other, with the latter the dominant one which makes neutron heavier. However in our dark baryonic sector the two can be additive. The radiative self energy correction will contribute a fixed mass splitting of about $0.1\%$ of the baryon mass, which alone will be sufficient for a light $m_{W_R}<10^8$~GeV or a low seesaw scale, nothing else is needed.

%%%%%%%%%%%%%%%%%%%%%%%%%%%%%%%%%%%%%%%%%%%%%%%%%%%%%%%%%%
\begin{figure}[t!]
\centering
\includegraphics[width=3.5in]{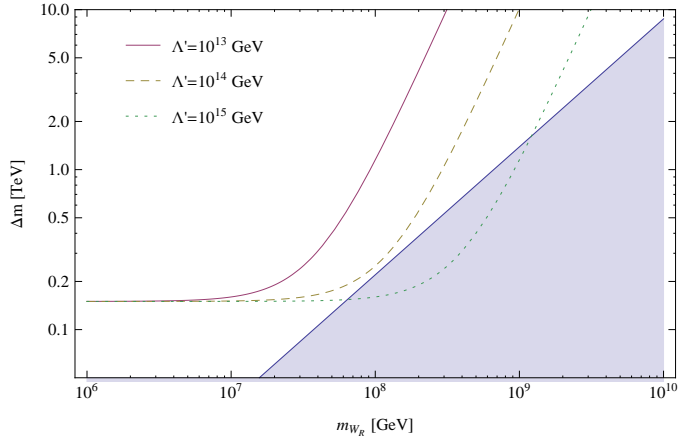}
\caption{The big bang nucleosynthesis constraint of $\tau\lesssim1$~s (shaded region excluded), on the dark neutron proton mass splitting $\Delta m=0.15~\text{TeV}+m_{W_R}^2/\Lambda'$ (electromagnetic radiative self energy correction plus current quark mass splitting from dimension-5 operator) with $m_{W_R}$.}
\label{fig:scales}
\end{figure}
%%%%%%%%%%%%%%%%%%%%%%%%%%%%%%%%%%%%%%%%%%%%%%%%%%%%%%%%%%

On the other hand, for a heavy $m_{W_R}$ the radiative self energy correction is not enough, and the current quark mass splitting shall be the dominant source of mass differences between dark neutron and proton. In our setup it can be generated by the dimension-5 operator $\bar{Q}\Phi^\dag\Phi Q$, in which the vacuum expectation value of $\Phi$ indeed makes the ``up type'' dark quark as well as the dark neutron heavier (\emph{viz.}~we give a theory for the up-down current quark mass splitting which is unexplained in QCD). Assuming the $SU(2)_R$ coupling is order one, the vacuum expectation value of $\Phi$ is expected to be at the same order of $m_{W_R}$. Here we need to introduce a new effective suppression scale $\Lambda'$ for the dimension-5 operator. In Fig.~\ref{fig:scales} we plot the $m_{W_R}$ vs.~$\Delta m$ with several $\Lambda'$s. We can see that the big bang nucleosynthesis constraint only exclude some region of $\Lambda'\gtrsim10^{14}$~GeV, even when the $\Delta m$ is reasonably small.

\subsection{The Dark QCD Spectrum}

In addition to the baryonic new particles in the dark sector possibly there are also mesonic new particles. Among them the most interesting ones will be the lightest, which are the pseudo Nambu-Goldstone (pNG) bosons\footnote{We assume some other broken symmetry to protect this Nambu-Goldstone bosons from having large masses. Note that one cannot equate the $SU(2)_R$ as the isospin $SU(2)_L$ in real QCD the chiral breaking of which leads to the QCD pion, for the quarks in our scaled-up QCD are vector-like.}. Corresponding to the QCD pions in the hypothetical proton decay, they can be the other decay product of the dark neutron except for the right hand charged leptons. Like the SM pions they are not stable either, the charged one can decay through a virtual $W_R$ into SM fermions, with the chirality flipping mechanism making the top bottom pair the dominant channel, and the neutral one will dominantly decay into photon pairs through anomalous triangle  diagram.

In real QCD the pion mass is determined by the Gell-Mann-Oakes-Renner relation~\cite{GellMann:1968rz} from the partially conserved axial current principle
\begin{equation}
m_\pi^2=-\frac{(m_u+m_d)}{f_\pi^2}\langle \bar{q} q\rangle
\end{equation}
In our dark QCD with the baryonic particle scale of $150$~TeV, we expect the $-\langle \bar{q} q\rangle/f_\pi^2$ also at the scale of hundred TeVs. The current quark mass is a complete free parameter, making the dark pNG pion mass also a free parameter. On the experimental side, the charged component of the dark pion should decay through a virtual $W_R$ which is exactly like the QCD pion decay through a virtual SM $W$, and chirality flipping mechanism makes the $t_R\bar{b}_R$ channel the dominant one. However the usual searches for $W'^\pm\to tb$ (or the $H^\pm\to tb$ which shares the same final state) assumes a singly produced $\Pi^\pm$, and in our model the dark pion should be produced in pairs by a very energetic gamma. Since searches for the top and bottom dijets in pairs have not been performed in this topology, and the production is through electroweak process which is very suppressed compared to a QCD process, there is essentially no experimental constraint.

We also expect other mesons in the spectrum which differ in discrete quantum numbers, such as the dark $\eta$ meson, the $\rho$ meson, the $\omega$ meson and so on. Without a universal understanding of the spectrum in strong dynamics we do not know the exact mass comparison with the dark neutron, so we will ignore their contribution in the dark neutron decay.

%%%%%%%%%%%%%%%%%%%%%%%%%%%%%%%
%%%%%%%%%%% Phenomenology %%%%%%%%%%%
%%%%%%%%%%%%%%%%%%%%%%%%%%%%%%%
\section{DM Decay Detectibility\label{sec:dec}}

Next we will compare the theoretical prediction of the $\mathcal{O}(10^{27}~s)$ DM decay life time to current and future experiments, to see whether they are detectable. We find that the AMS-02 preliminary antiproton data indeed favors an $\mathcal{O}(10^{27}~s)$ DM decay life time, and it does not exceed the Fermi EGB bound and the IceCube neutrino bound. As a benchmark we ignore all the SM particle mass including the top quark which are much smaller, and choose the dark pion mass to be $1/10$ of the dark baryon mass or $15$~TeV. We want to emphasize that it should not be taken as a rigorous data driven fit.

Except for the three DM indirect detections, we found direct detection experiments are also constraining and complementary.
Our result agrees with~\cite{Antipin:2015xia} and we would like to redirect reader to their paper but not repeat it here. %However, direct detection experiments are not able to constrain on DM decay time.

\subsection{Proton and Antiproton ratio}

The dark neutron will decay through the proton decay like chain of $N\to \ell^+_R\Pi^-\to \ell^+_R\bar{t}_Rb_R$ as discussed before. The neutral decay mode of replacing the $\ell^+_R$ by the (on-shell) right hand antineutrino is kinematically forbidden. The branching ratio for $\ell^+$ is assumed equal for three families. The same but every particle replaced by antiparticle decay chain applies to dark antineutron.

For the most interesting decay product of top quark, the energy distribution can be determined analytically in the sequential two body decay. With $m_\Pi=15~\text{TeV}$ fixed the charged lepton has a fixed energy of $E_\ell=(m_\chi^2-m_\Pi^2)/(2m_\chi)=74.25$~TeV, and the top and bottom will be evenly distributed in the energy region from $E_{q\text{min}}=\frac{1}{2}(E_\Pi-\sqrt{E_\Pi^2-m_\Pi^2})$ to $E_{q\text{max}}=\frac{1}{2}(E_\Pi+\sqrt{E_\Pi^2-m_\Pi^2})$, where $E_\Pi=(m_\chi^2+m_\Pi^2)/(2m_\chi)=75.75$~TeV. All the prompt decay spectra are then calculated by
\begin{align}
\frac{dN_i}{dE}&=\sum_{q=t,b}\int_{E_{q\text{min}}}^{E_{q\text{max}}}
\frac{dE_q}{\sqrt{E_\Pi^2-m_\Pi^2}}\bigg(\frac{dN_i}{dE}\bigg)'_q\big(m_\chi=E_q\big)\nonumber\\
&+\frac{1}{3}\sum_{\ell=e,\mu,\tau}\bigg(\frac{dN_i}{dE}\bigg)'_\ell\big(m_\chi=E_\ell\big)~,\label{eq:prompt}
\end{align}
where the primed $\frac{dN_i}{dE}$ are taken from the \texttt{PPPC4}~\cite{Cirelli:2010xx} for $i=p\bar{p},e^+e^-,\gamma$ and so on. \footnote{The \texttt{PPPC4} cosmic ray spectrum calculation is only available for unpolarized primary quarks, but in our model the primary $\bar{t}_R$ and $b_R$ are all right-handed. In parton shower and hadronization process the helicity induced difference in yield spectrum is arguably small (which can affect a few percent only in the high energy region), hence we directly use the unpolarized result.}

\begin{figure}[t!]
\centering
\includegraphics[width=3in]{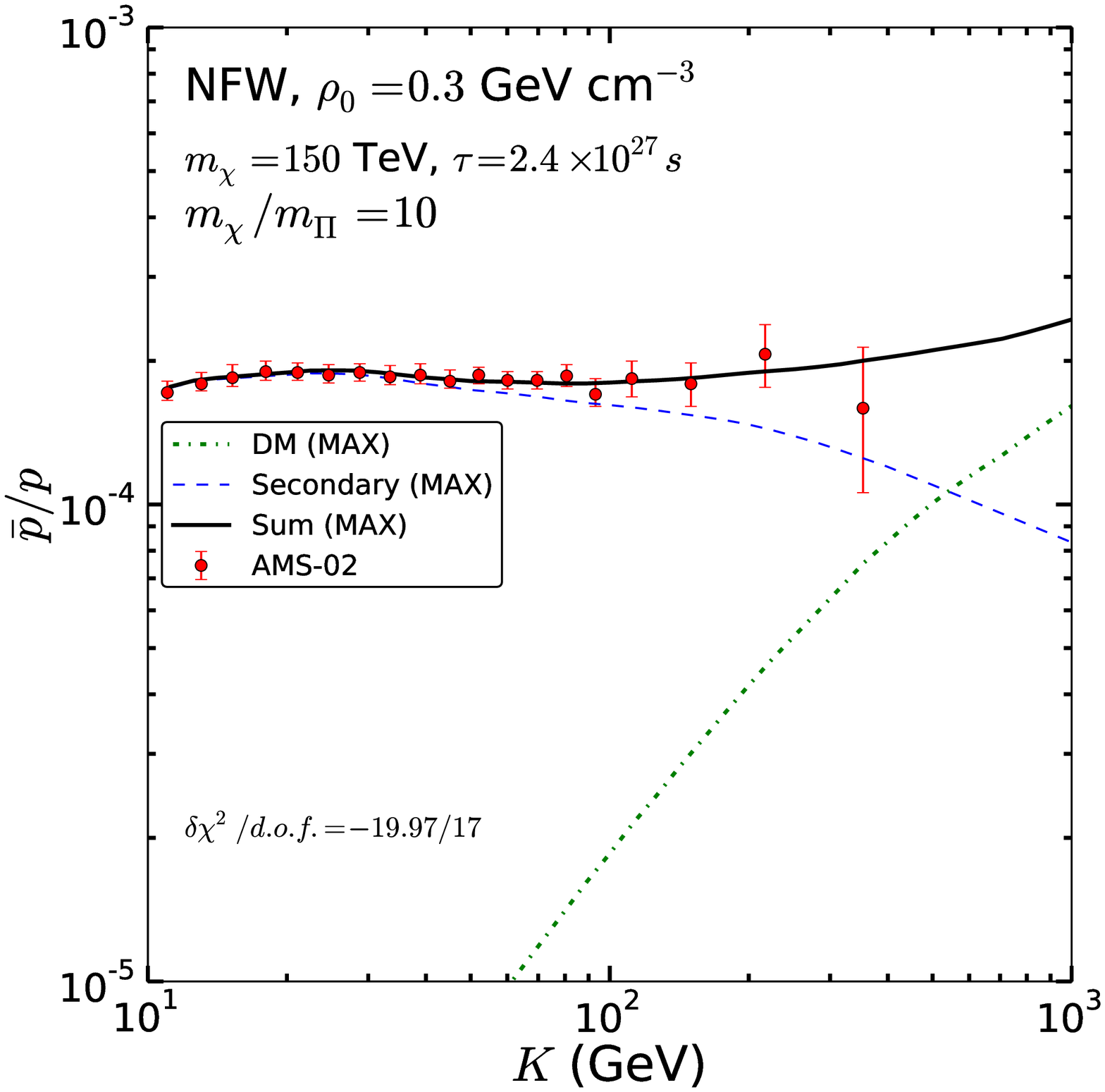}~~
\includegraphics[width=3in]{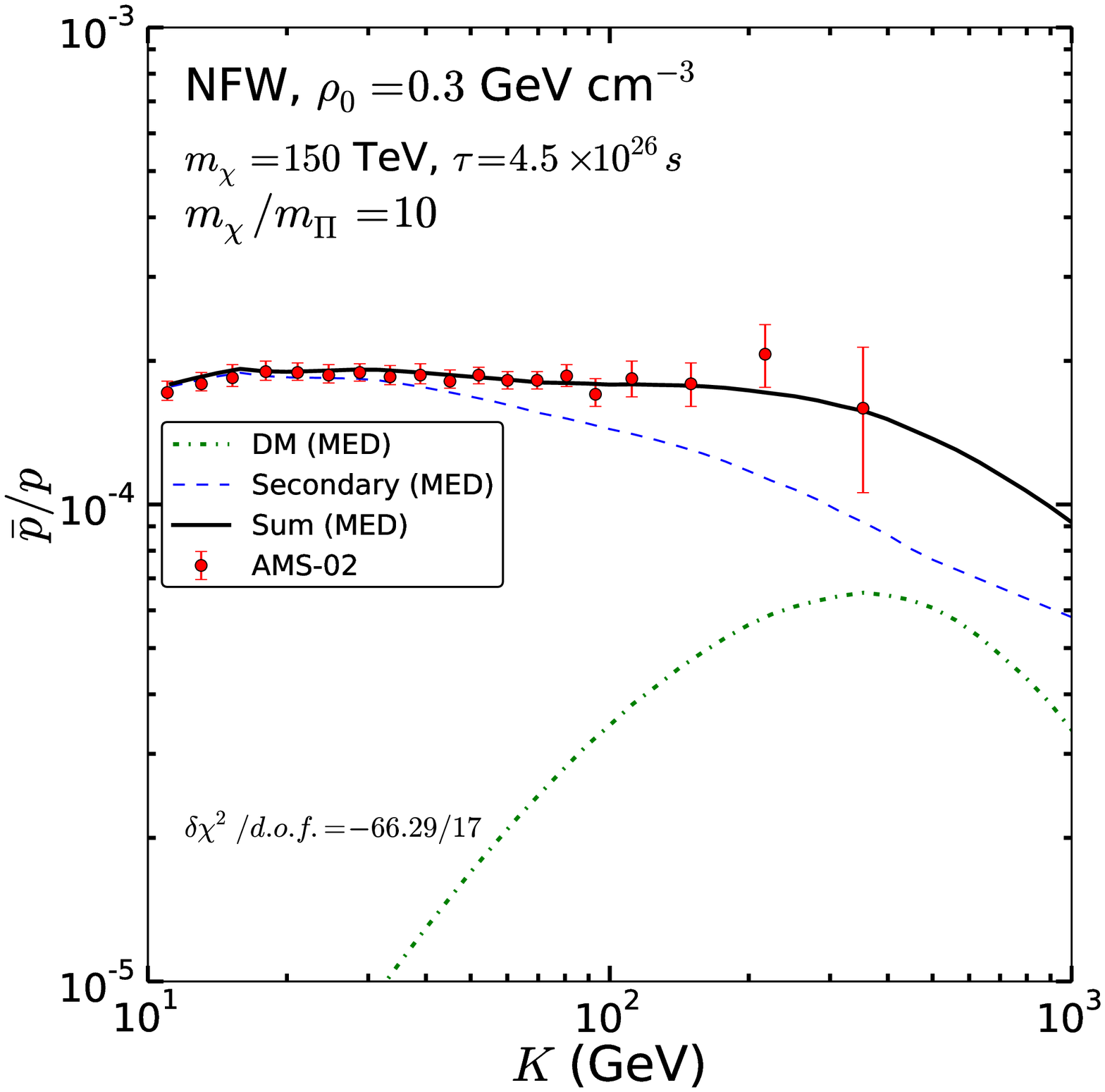}
\caption{The antiproton to proton ratio fitting to the AMS-02 preliminary data for the \texttt{MAX} (left panel) and \texttt{MED} (right panel) propagation parameters. The DM alone, background alone, and sum of DM and background contributions are shown as green dashed-dotted line, blue dashed line and the black line respectively.}
\label{fig:proton}
\end{figure}

The antiproton prediction are shown in Fig.~\ref{fig:proton}, to the AMS-02 preliminary antiproton data~\cite{AMS-02} which shows a slight excess but still can be viewed as consistent with background. In this work we use the the NFW halo profile and the \texttt{MAX}/\texttt{MED} propagation parameters~\cite{Donato:2003xg,Delahaye:2007fr} for antiproton, as argued recently in~\cite{Kappl:2015bqa} that the \texttt{MED} is less favored by isotopic abundance data from AMS-02, but for completeness we still leave our prediction for \texttt{MED} on the right panel.
The background is taken from~\cite{Giesen:2015ufa}. Considering $K>10$~GeV, the best-fit DM decay time is $\tau=2.4\times 10^{27}$ second for \texttt{MAX} and $\tau=4.5\times 10^{26}$ second for \texttt{MED}. The chi-square improvement of $(\chi^2_\text{Sum}-\chi^2_\text{Secondary})/d.o.f.$ can achieve $19.97/17$ ($66.29/17$) for \texttt{MAX} (\texttt{MED}).

We have checked that varying the dark pion mass (e.g., by a factor of two) has negligible effect on the spectrum (Eq.~\ref{eq:prompt}), as well as the experimental statistical strength. The result always holds as long as the dark pion mass is much smaller than the dark matter mass.

\subsection{Extragalactic Gamma Ray and Neutrino}

\begin{figure}[t!]
\centering
\includegraphics[width=3in]{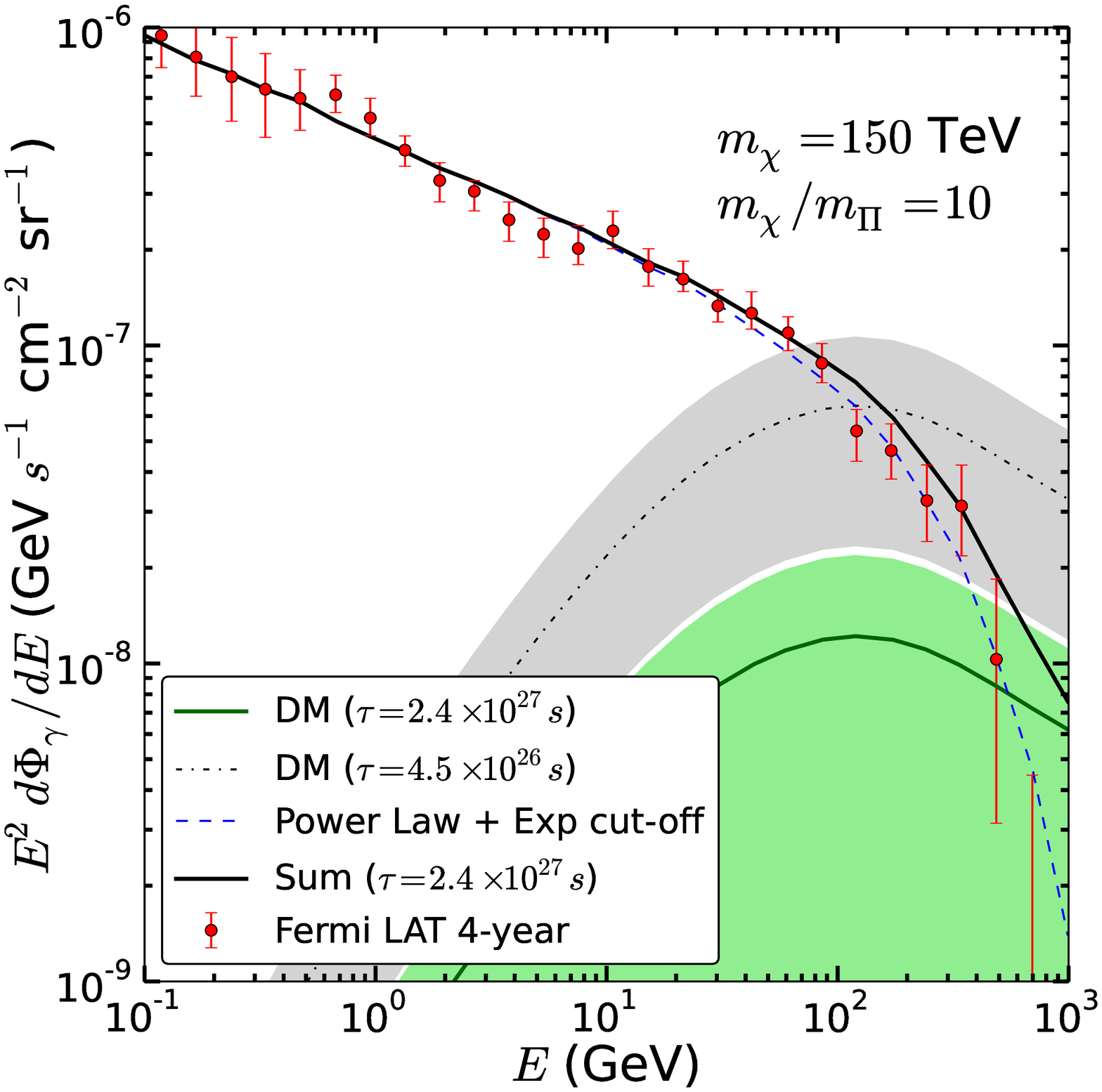}~~
\includegraphics[width=3in]{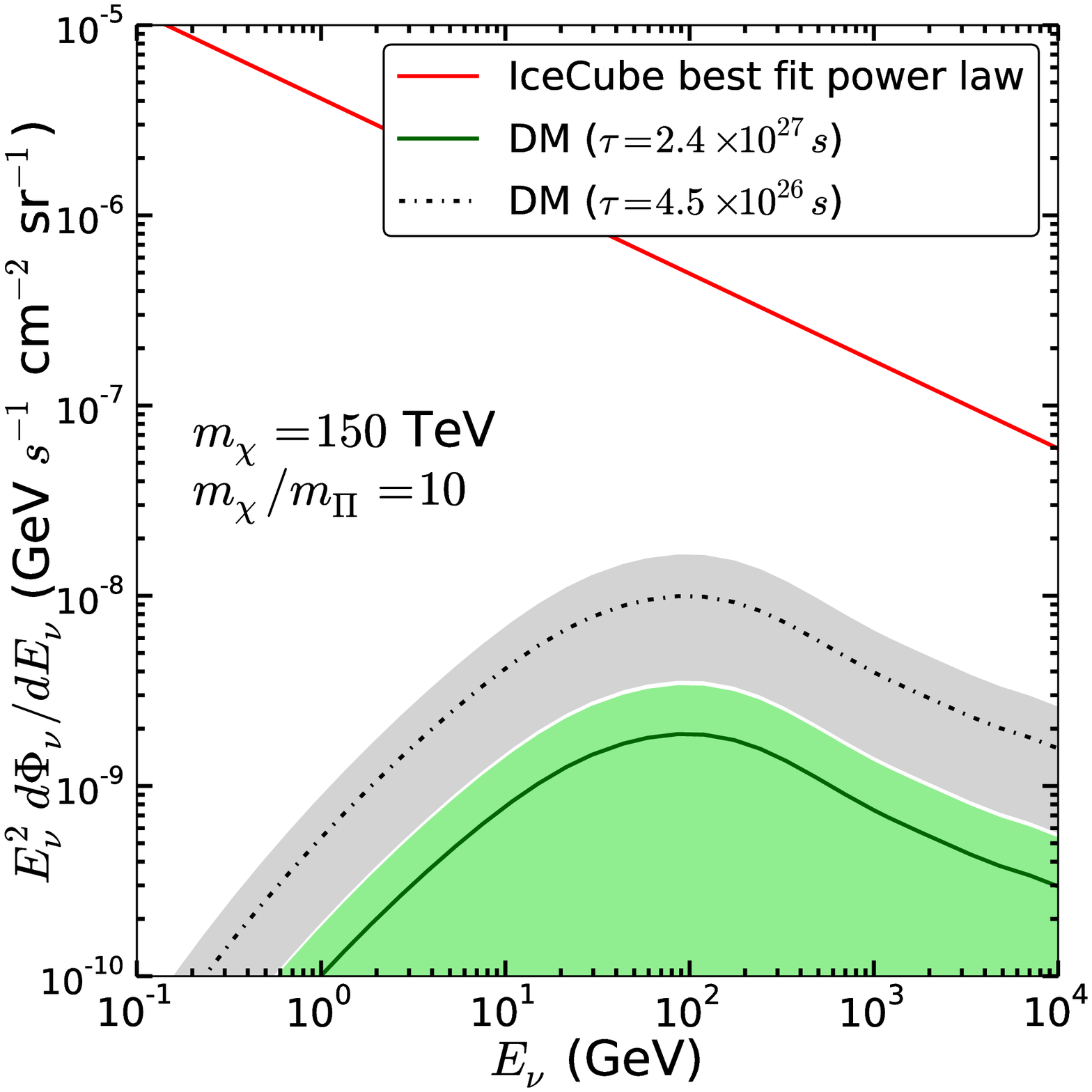}
\caption{
The EGB gamma ray (left panel) and neutrino (right panel) energy spectrum predictions based on two benchmark decay times. The gray and green bands represent the $2\sigma$ favored decay time by AMS-02 antiproton data with propagation \texttt{MED} and \texttt{MAX} respectively.}
\label{fig:gamma}
\end{figure}
The EGB\footnote{In~\cite{Regis:2015zka} it is pointed out that the most stringent gamma ray limit for decaying DM is from the angular cross-correlation of low-redshift sources, but the improvement, c.f.~\cite{Ando:2015qda}, seems weaker in our high energy region.} prediction comparing with the Fermi-LAT 4 year data~\cite{Ackermann:2015tah} are shown in Fig.~\ref{fig:gamma}.
We ignored the local DM decay contribution such as that from our Milky Way, only working for redshift $z_\text{min}=10^{-4}$ to $z_\text{max}=2$ where the gamma ray is effectively cut off by the optical depth for all interested energy. We also ignore the contribution from inverse Compton scattering of our charged decay product with the CMB photons, which is less important in our interested high energy region and makes the result conservative. The gamma ray flux can be described by
\begin{equation}
\frac{d\Phi_{\gamma}}{dE}=\frac{c}{4\pi}\frac{\Omega_\chi\rho_c}{m_\chi\tau_{\chi}} \int_{z_\text{min}}^{z_\text{max}}dz\frac{e^{-\tau(z,E)}}{H(z)}\frac{dN_\gamma}{dE}(E^{z})~,\label{eq:EGB}
\end{equation}
where $H(z)\simeq H_0\sqrt{(\Omega_\chi+\Omega_b)(1+z)^3+\Omega_{\Lambda}}$ is the Hubble function and the data of $H_0,\Omega_\chi,\Omega_b,\Omega_\Lambda$ and critical density $\rho_c$ are taken from~\cite{Ade:2013zuv}. The redshifted photon energy $E$ measured at earth is related to the initial energy of $E^{z}=E(1+z)$ at production, and the optical depth suppression $e^{-\tau(z,E)}$ is taken of the min UV case in \texttt{PPPC4}~\cite{Cirelli:2010xx}. For the background we use a power law with cutoff, ignoring possible astrophysical sources such as blazars, star-forming galaxies~\cite{Ackermann:2012vca}, and misaligned active galactic nuclei~\cite{DiMauro:2013xta,Inoue:2011bm}.

In general we find an $\mathcal{O}(10^{27})~s$ decay life time is consistent with the Fermi EGB gamma ray data. For example, with $\tau=2.4\times10^{27}~s$ favored by the AMS-02 antiproton data with \texttt{MAX} propagation, our model prediction is well below the Fermi EGB gamma ray data. On the other hand, the $\tau=4.5\times 10^{26}~s$ favored by antiproton data with \texttt{MED} propagation is almost excluded, but still barely allowed if including the $2\sigma$ error band of the antiproton data.

Also the $\mathcal{O}(10^{27})~s$ decay life time is allowed in the neutrino channel. There is no recent isotropic neutrino data in our desired energy regime\footnote{In our model we cannot have primary SM neutrinos and all produced SM
neutrinos are secondary, and their energy should be much smaller than the
DM scale of 150 TeV, so unable to account the IceCube 100 TeV to PeV neutrino signals.}, so we use an extrapolation~\cite{Aartsen:2014muf} fitted from the IceCube neutrino data at an even higher energy spectrum~\cite{Aartsen:2014gkd}, assuming a single unbroken power law and equal neutrino fluxes of all flavors. In the right panel of Fig.~\ref{fig:gamma}, it is shown that the $\mathcal{O}(10^{27})~s$ decay life time is at least two orders below the current IceCube neutrino extrapolated bound.

%%%%%%%%%%%%%%%%%%%%%%%%%%%%%%%
%%%%%%%%%%% Summary %%%%%%%%%%%
%%%%%%%%%%%%%%%%%%%%%%%%%%%%%%%
\section{Summary\label{sec:sum}}

We have proposed a new scenario of baryonic DM based on a strong hidden $SU(3)$ gauge group, which also connects to the visible sector through the $SU(2)_R\times U(1)_{B-L}$ gauge group. This composite gauge group is primarily motivated by the DM stability consideration, but the hypothetical Planck scale suppressed decay can induce interesting possible signals while the connection sector $SU(2)_R\times U(1)_{B-L}$ is added.

In the minimal model building we introduce the following scales
\begin{itemize}
\item The DM or the dark baryon mass scale of $150$~TeV, which is determined by rescaling the real QCD baryon antibaryon annihilation cross section to the ``WIMP miracle'' cross section.
\item The dark pion mass scale which is much lower.
\item The $SU(2)_R$ breaking scale, which is free but appears in the natural seesaw mechanism.
\item The optional free effective suppression scale of the dimension-5 operator.
\end{itemize}

We find that theoretical estimation and fitting to recent preliminary AMS-02 antiproton data are both consistent with a decay life time of a few $10^{27}~s$. Although being heavy, our DM model has a perspective of detection in the near future.

%%%%%%%%%%%%%%%%%%%%%%%%%%%%%%%%%%%%%%
%%%%%%%%%%% Acknowledgement %%%%%%%%%%%
%%%%%%%%%%%%%%%%%%%%%%%%%%%%%%%%%%%%%%
\bigskip
{\bf Acknowledgments}
\vspace{0.1cm}\\
\noindent
The authors are grateful to Yoshimasa Hidaka, Masahiro Maruyama, Yu-Feng Zhou and Shin'ichiro Ando for useful discussions. This work is supported by the Grant-in-Aid for Scientific research from the Ministry of Education, Science, Sports, and Culture (MEXT), Japan [No. 26104009/26287039 and 16H02176 (S.M. and T.T.Y.)], as well as by the World Premier International Research Center Initiative (WPI), MEXT, Japan.

%%%%%%%%%%%%%%%%%%%%%%%%%%%%%%%%%
%%%%%%%%%%% References %%%%%%%%%%%
%%%%%%%%%%%%%%%%%%%%%%%%%%%%%%%%%
%\input{references}


\begin{thebibliography}{99}

%\cite{Ade:2015xua}
\bibitem{Ade:2015xua}
  P.~A.~R.~Ade {\it et al.} [Planck Collaboration],
  %``Planck 2015 results. XIII. Cosmological parameters,''
  arXiv:1502.01589 [astro-ph.CO].
  %%CITATION = ARXIV:1502.01589;%%
  %1119 citations counted in INSPIRE as of 18 Jan 2016

%\cite{Nussinov:1985xr}
\bibitem{Nussinov:1985xr}
  S.~Nussinov,
  %``Technocosmology: Could A Technibaryon Excess Provide A 'natural' Missing Mass Candidate?,''
  Phys.\ Lett.\ B {\bf 165}, 55 (1985).
  %%CITATION = PHLTA,B165,55;%%
  %275 citations counted in INSPIRE as of 28 Oct 2015

%\cite{Chivukula:1989qb}
\bibitem{Chivukula:1989qb}
  R.~S.~Chivukula and T.~P.~Walker,
  %``Technicolor Cosmology,''
  Nucl.\ Phys.\ B {\bf 329}, 445 (1990).
  %%CITATION = NUPHA,B329,445;%%
  %95 citations counted in INSPIRE as of 28 Oct 2015

%\cite{Barr:1990ca}
\bibitem{Barr:1990ca}
  S.~M.~Barr, R.~S.~Chivukula and E.~Farhi,
  %``Electroweak Fermion Number Violation and the Production of Stable Particles in the Early Universe,''
  Phys.\ Lett.\ B {\bf 241}, 387 (1990).
  %%CITATION = PHLTA,B241,387;%%
  %221 citations counted in INSPIRE as of 28 Oct 2015

%\cite{Gudnason:2006yj}
\bibitem{Gudnason:2006yj}
  S.~B.~Gudnason, C.~Kouvaris and F.~Sannino,
  %``Dark Matter from new Technicolor Theories,''
  Phys.\ Rev.\ D {\bf 74}, 095008 (2006)
  %doi:10.1103/PhysRevD.74.095008
  [hep-ph/0608055].
  %%CITATION = doi:10.1103/PhysRevD.74.095008;%%
  %177 citations counted in INSPIRE as of 21 Jul 2016

%\cite{Murayama:2009nj}
\bibitem{Murayama:2009nj}
  H.~Murayama and J.~Shu,
  %``Topological Dark Matter,''
  Phys.\ Lett.\ B {\bf 686}, 162 (2010)
  [arXiv:0905.1720 [hep-ph]].
  %%CITATION = ARXIV:0905.1720;%%
  %24 citations counted in INSPIRE as of 19 Jul 2015

%\cite{Hambye:2009fg}
\bibitem{Hambye:2009fg}
  T.~Hambye and M.~H.~G.~Tytgat,
  %``Confined hidden vector dark matter,''
  Phys.\ Lett.\ B {\bf 683}, 39 (2010)
  %doi:10.1016/j.physletb.2009.11.050
  [arXiv:0907.1007 [hep-ph]].
  %%CITATION = doi:10.1016/j.physletb.2009.11.050;%%
  %49 citations counted in INSPIRE as of 17 Aug 2016

%\cite{Hamaguchi:2009db}
\bibitem{Hamaguchi:2009db}
  K.~Hamaguchi, E.~Nakamura, S.~Shirai and T.~T.~Yanagida,
  %``Low-Scale Gauge Mediation and Composite Messenger Dark Matter,''
  JHEP {\bf 1004}, 119 (2010)
  [arXiv:0912.1683 [hep-ph]].
  %%CITATION = ARXIV:0912.1683;%%
  %6 citations counted in INSPIRE as of 04 Mar 2015

%\cite{Antipin:2014qva}
\bibitem{Antipin:2014qva}
  O.~Antipin, M.~Redi and A.~Strumia,
  %``Dynamical generation of the weak and Dark Matter scales from strong interactions,''
  JHEP {\bf 1501}, 157 (2015)
  [arXiv:1410.1817 [hep-ph]].
  %%CITATION = ARXIV:1410.1817;%%
  %12 citations counted in INSPIRE as of 02 Jul 2015

%\cite{Antipin:2015xia}
\bibitem{Antipin:2015xia}
  O.~Antipin, M.~Redi, A.~Strumia and E.~Vigiani,
  %``Accidental Composite Dark Matter,''
  JHEP {\bf 1507}, 039 (2015)
  %doi:10.1007/JHEP07(2015)039
  [arXiv:1503.08749 [hep-ph]].
  %%CITATION = doi:10.1007/JHEP07(2015)039;%%
  %22 citations counted in INSPIRE as of 19 Aug 2016

%\cite{Ibarra:2015tya}
\bibitem{Ibarra:2015tya}
  A.~Ibarra, A.~S.~Lamperstorfer, S.~L\'{o}pez-Gehler, M.~Pato and G.~Bertone,
  %``On the sensitivity of CTA to gamma-ray boxes from multi-TeV dark matter,''
  JCAP {\bf 1509}, no. 09, 048 (2015)
  Erratum: [JCAP {\bf 1606}, no. 06, E02 (2016)]
  %doi:10.1088/1475-7516/2016/06/E02, 10.1088/1475-7516/2015/09/048
  [arXiv:1503.06797 [astro-ph.HE]].
  %%CITATION = doi:10.1088/1475-7516/2016/06/E02, 10.1088/1475-7516/2015/09/048;%%
  %21 citations counted in INSPIRE as of 19 Aug 2016

%\cite{Griest:1989wd}
\bibitem{Griest:1989wd}
  K.~Griest and M.~Kamionkowski,
  %``Unitarity Limits on the Mass and Radius of Dark Matter Particles,''
  Phys.\ Rev.\ Lett.\  {\bf 64}, 615 (1990).
  %%CITATION = PRLTA,64,615;%%
  %254 citations counted in INSPIRE as of 17 Jul 2014

%\cite{Hisano:2003ec}
\bibitem{Hisano:2003ec}
  J.~Hisano, S.~Matsumoto and M.~M.~Nojiri,
  %``Explosive dark matter annihilation,''
  Phys.\ Rev.\ Lett.\  {\bf 92}, 031303 (2004);
  %[hep-ph/0307216].
  %%CITATION = HEP-PH/0307216;%%
  %231 citations counted in INSPIRE as of 14 Dec 2014

%\cite{Hisano:2004ds}
\bibitem{Hisano:2004ds}
  J.~Hisano, S.~Matsumoto, M.~M.~Nojiri and O.~Saito,
  %``Non-perturbative effect on dark matter annihilation and gamma ray signature from galactic center,''
  Phys.\ Rev.\ D {\bf 71}, 063528 (2005).
  [hep-ph/0412403].
  %%CITATION = HEP-PH/0412403;%%
  %279 citations counted in INSPIRE as of 10 Dec 2014

%\cite{Bruckner:1989ew}
\bibitem{Bruckner:1989ew}
  W.~Bruckner, B.~Cujec, H.~Dobbeling, K.~Dworschak, F.~Guttner, H.~Kneis, S.~Majewski and M.~Nomachi {\it et al.},
  %``Measurements of the Anti-proton - Proton Annihilation Cross-section in the Beam Momentum Range Between 180-{MeV}/$c$ and 600-{MeV}/$c$,''
  Z.\ Phys.\ A {\bf 335}, 217 (1990).
  %%CITATION = ZEPYA,A335,217;%%
  %33 citations counted in INSPIRE as of 16 Jun 2015

%\cite{Bertin:1996kw}
\bibitem{Bertin:1996kw}
  A.~Bertin {\it et al.}  [OBELIX Collaboration],
  %``anti-p p annihilation cross-section at very low-energy,''
  Phys.\ Lett.\ B {\bf 369}, 77 (1996).
  %%CITATION = PHLTA,B369,77;%%
  %27 citations counted in INSPIRE as of 16 Jun 2015

%\cite{Benedettini:1997fk}
\bibitem{Benedettini:1997fk}
  A.~Benedettini {\it et al.}  [OBELIX Collaboration],
  %``anti-p p partial cross-sections at low energy,''
  Nucl.\ Phys.\ Proc.\ Suppl.\  {\bf 56}, 58 (1997).
  %%CITATION = NUPHZ,56,58;%%
  %7 citations counted in INSPIRE as of 16 Jun 2015

%\cite{Zenoni:1999st}
\bibitem{Zenoni:1999st}
  A.~Zenoni, A.~Bianconi, F.~Bocci, G.~Bonomi, M.~Corradini, A.~Donzella, E.~Lodi Rizzini and L.~Venturelli {\it et al.},
  %``New measurements of the anti-p p annihilation cross-section at very low-energy,''
  Phys.\ Lett.\ B {\bf 461}, 405 (1999).
  %%CITATION = PHLTA,B461,405;%%
  %28 citations counted in INSPIRE as of 16 Jun 2015

%\cite{Armstrong:1987nu}
\bibitem{Armstrong:1987nu}
  T.~Armstrong {\it et al.}  [BROOKHAVEN-HOUSTON-PENNSYLVANIA STATE-RICE Collaboration],
  %``Measurement of Anti-neutron Proton Total and Annihilation Cross-sections From 100-{MeV}/c to 500-{MeV}/c,''
  Phys.\ Rev.\ D {\bf 36}, 659 (1987).
  %%CITATION = PHRVA,D36,659;%%
  %37 citations counted in INSPIRE as of 16 Jun 2015

%\cite{Bertin:1997gn}
\bibitem{Bertin:1997gn}
  A.~Bertin {\it et al.}  [OBELIX Collaboration],
  %``anti-n p annihilation in flight in two mesons in the momentum range between 50-MeV/c and 400-MeV/c with OBELIX,''
  Nucl.\ Phys.\ Proc.\ Suppl.\  {\bf 56}, 227 (1997).
  %%CITATION = NUPHZ,56,227;%%
  %4 citations counted in INSPIRE as of 16 Jun 201

%\cite{Aoki:2013yxa}
\bibitem{Aoki:2013yxa}
  Y.~Aoki, E.~Shintani and A.~Soni,
  %``Proton decay matrix elements on the lattice,''
  Phys.\ Rev.\ D {\bf 89}, no. 1, 014505 (2014)
  [arXiv:1304.7424 [hep-lat]].
  %%CITATION = ARXIV:1304.7424;%%
  %22 citations counted in INSPIRE as of 18 juin 2015

\bibitem{neutrinobook}
  M.~Fukugita, T.~T.~Yanagida,
  ``Physics of Neutrinos: And Applications to Astrophysics''.

%\cite{Wilkinson:1982hu}
\bibitem{Wilkinson:1982hu}
  D.~H.~Wilkinson,
  %``Analysis Of Neutron Beta Decay,''
  Nucl.\ Phys.\ A {\bf 377}, 474 (1982).
  %%CITATION = NUPHA,A377,474;%%
  %151 citations counted in INSPIRE as of 18 Jun 2015

%\cite{GellMann:1968rz}
\bibitem{GellMann:1968rz}
  M.~Gell-Mann, R.~J.~Oakes and B.~Renner,
  %``Behavior of current divergences under SU(3) x SU(3),''
  Phys.\ Rev.\  {\bf 175}, 2195 (1968).
  %%CITATION = PHRVA,175,2195;%%
  %1554 citations counted in INSPIRE as of 23 Jun 2015

%\cite{Cirelli:2010xx}
\bibitem{Cirelli:2010xx}
  M.~Cirelli, G.~Corcella, A.~Hektor, G.~Hutsi, M.~Kadastik, P.~Panci, M.~Raidal and F.~Sala {\it et al.},
  %``PPPC 4 DM ID: A Poor Particle Physicist Cookbook for Dark Matter Indirect Detection,''
  JCAP {\bf 1103}, 051 (2011)
  [JCAP {\bf 1210}, E01 (2012)]
  [arXiv:1012.4515 [hep-ph], arXiv:1012.4515 [hep-ph]].
  %%CITATION = ARXIV:1012.4515;%%
  %214 citations counted in INSPIRE as of 17 juin 2015

%\cite{AMS-02}
\bibitem{AMS-02}
  AMS-02 Collaboration, Talks at the `AMS Days at CERN', 15-17, April, 2015.

%\cite{Donato:2003xg}
\bibitem{Donato:2003xg}
  F.~Donato, N.~Fornengo, D.~Maurin, P.~Salati and R.~Taillet,
  %``Antiprotons in cosmic rays from neutralino annihilation,''
  Phys.\ Rev.\ D {\bf 69}, 063501 (2004)
  [astro-ph/0306207].
  %%CITATION = ASTRO-PH/0306207;%%
  %243 citations counted in INSPIRE as of 17 juin 2015

%\cite{Delahaye:2007fr}
\bibitem{Delahaye:2007fr}
  T.~Delahaye, R.~Lineros, F.~Donato, N.~Fornengo and P.~Salati,
  %``Positrons from dark matter annihilation in the galactic halo: Theoretical uncertainties,''
  Phys.\ Rev.\ D {\bf 77}, 063527 (2008)
  [arXiv:0712.2312 [astro-ph]].
  %%CITATION = ARXIV:0712.2312;%%
  %215 citations counted in INSPIRE as of 17 Jun 2015

%\cite{Kappl:2015bqa}
\bibitem{Kappl:2015bqa}
  R.~Kappl, A.~Reinert and M.~W.~Winkler,
  %``AMS-02 Antiprotons Reloaded,''
  arXiv:1506.04145 [astro-ph.HE].
  %%CITATION = ARXIV:1506.04145;%%

%\cite{Giesen:2015ufa}
\bibitem{Giesen:2015ufa}
  G.~Giesen, M.~Boudaud, Y.~Genolini, V.~Poulin, M.~Cirelli, P.~Salati and P.~D.~Serpico,
  %``AMS-02 antiprotons, at last! Secondary astrophysical component and immediate implications for Dark Matter,''
  arXiv:1504.04276 [astro-ph.HE].
  %%CITATION = ARXIV:1504.04276;%%
  %18 citations counted in INSPIRE as of 15 juin 2015

%\cite{Regis:2015zka}
\bibitem{Regis:2015zka}
  M.~Regis, J.~Q.~Xia, A.~Cuoco, E.~Branchini, N.~Fornengo and M.~Viel,
  %``Particle dark matter searches outside the Local Group,''
  arXiv:1503.05922 [astro-ph.CO].
  %%CITATION = ARXIV:1503.05922;%%
  %1 citations counted in INSPIRE as of 17 juin 2015

%\cite{Ando:2015qda}
\bibitem{Ando:2015qda}
  S.~Ando and K.~Ishiwata,
  %``Constraints on decaying dark matter from the extragalactic gamma-ray background,''
  JCAP {\bf 1505}, no. 05, 024 (2015)
  [arXiv:1502.02007 [astro-ph.CO]].
  %%CITATION = ARXIV:1502.02007;%%
  %7 citations counted in INSPIRE as of 16 Jun 2015

%\cite{Ackermann:2015tah}
\bibitem{Ackermann:2015tah}
  M.~Ackermann {\it et al.}  [Fermi-LAT Collaboration],
  %``Limits on Dark Matter Annihilation Signals from the Fermi LAT 4-year Measurement of the Isotropic Gamma-Ray Background,''
  arXiv:1501.05464 [astro-ph.CO].
  %%CITATION = ARXIV:1501.05464;%%
  %16 citations counted in INSPIRE as of 15 juin 2015

%\cite{Ade:2013zuv}
\bibitem{Ade:2013zuv}
  P.~A.~R.~Ade {\it et al.} [Planck Collaboration],
  %``Planck 2013 results. XVI. Cosmological parameters,''
  Astron.\ Astrophys.\  {\bf 571}, A16 (2014)
  doi:10.1051/0004-6361/201321591
  [arXiv:1303.5076 [astro-ph.CO]].
  %%CITATION = doi:10.1051/0004-6361/201321591;%%
  %4267 citations counted in INSPIRE as of 26 Jan 2016

%\cite{Ackermann:2012vca}
\bibitem{Ackermann:2012vca}
  M.~Ackermann {\it et al.}  [Fermi-LAT Collaboration],
  %``GeV Observations of Star-forming Galaxies with \textit{Fermi} LAT,''
  Astrophys.\ J.\  {\bf 755}, 164 (2012)
  [arXiv:1206.1346 [astro-ph.HE]].
  %%CITATION = ARXIV:1206.1346;%%
  %74 citations counted in INSPIRE as of 16 juin 2015

%\cite{DiMauro:2013xta}
\bibitem{DiMauro:2013xta}
  M.~Di Mauro, F.~Calore, F.~Donato, M.~Ajello and L.~Latronico,
  %``Diffuse $\gamma$-ray emission from misaligned active galactic nuclei,''
  Astrophys.\ J.\  {\bf 780}, 161 (2014)
  [arXiv:1304.0908 [astro-ph.HE]].
  %%CITATION = ARXIV:1304.0908;%%
  %30 citations counted in INSPIRE as of 16 juin 2015

%\cite{Inoue:2011bm}
\bibitem{Inoue:2011bm}
  Y.~Inoue,
  %``Contribution of the Gamma-ray Loud Radio Galaxies Core Emissions to the Cosmic MeV and GeV Gamma-Ray Background Radiation,''
  Astrophys.\ J.\  {\bf 733}, 66 (2011)
  [arXiv:1103.3946 [astro-ph.HE]].
  %%CITATION = ARXIV:1103.3946;%%
  %59 citations counted in INSPIRE as of 16 Jun 2015

%\cite{Aartsen:2014muf}
\bibitem{Aartsen:2014muf}
  M.~G.~Aartsen {\it et al.} [IceCube Collaboration],
  %``Atmospheric and astrophysical neutrinos above 1 TeV interacting in IceCube,''
  Phys.\ Rev.\ D {\bf 91}, no. 2, 022001 (2015)
  doi:10.1103/PhysRevD.91.022001
  [arXiv:1410.1749 [astro-ph.HE]].
  %%CITATION = doi:10.1103/PhysRevD.91.022001;%%
  %91 citations counted in INSPIRE as of 26 janv. 2016

%\cite{Aartsen:2014gkd}
\bibitem{Aartsen:2014gkd}
  M.~G.~Aartsen {\it et al.} [IceCube Collaboration],
  %``Observation of High-Energy Astrophysical Neutrinos in Three Years of IceCube Data,''
  Phys.\ Rev.\ Lett.\  {\bf 113}, 101101 (2014)
  doi:10.1103/PhysRevLett.113.101101
  [arXiv:1405.5303 [astro-ph.HE]].
  %%CITATION = doi:10.1103/PhysRevLett.113.101101;%%
  %307 citations counted in INSPIRE as of 26 Jan 2016

\end{thebibliography}
\end{document}